\documentclass[aps,pre,twocolumn,showpacs,floatfix]{revtex4}
\usepackage{amssymb,amsmath,epsfig,graphicx}

\usepackage{caption}
\usepackage{dcolumn}

\newcommand{\vmedia}{\langle \dot x_{\text{cm}}\rangle}  
\newcommand{\vone}{\langle \dot x\rangle}     
\newcommand{\erfc}{\text{erfc}}
\newcommand{\Feff}{F_{\text{eff}}}
\newcommand{\Veff}{V_{\text{eff}}}
\newcommand{\ton}{t_{\text{on}}}
\newcommand{\toff}{t_{\text{off}}}

\newcommand{\Con}{C_{\text{on}}}
\newcommand{\Coff}{C_{\text{off}}}
\newcommand{\lambdaon}{\lambda_{\text{on}}}
\newcommand{\lambdaoff}{\lambda_{\text{off}}}

\newcommand{\taumin}{\tau_{\text{min}}}
\newcommand{\ierfc}{\text{ierfc}}
\newcommand{\sgn}{\text{sgn}}

\newcommand{\be}{\begin{equation}}
\newcommand{\ee}{\end{equation}}

\begin{document}

\title{\bf Time-delayed feedback control of a flashing ratchet}

\author{M. Feito}
\email{feito@fis.ucm.es}
\affiliation{Departamento de F\'{\i}sica At\'omica, Molecular y Nuclear,
Universidad Complutense de Madrid, \\
Avenida Complutense s/n, 28040 Madrid, Spain.}

\author{F. J. Cao}
\email{francao@fis.ucm.es}
\affiliation{Departamento de F\'{\i}sica At\'omica, Molecular y Nuclear,
Universidad Complutense de Madrid, \\
Avenida Complutense s/n, 28040 Madrid, Spain}
\affiliation{LERMA, Observatoire de Paris, Laboratoire Associ\'e au CNRS UMR 8112, \\
61, Avenue de l'Observatoire, 75014 Paris, France.}


\begin{abstract}     
Closed-loop or feedback control ratchets use information about the state 
of the system to operate with the aim of maximizing the performance of the 
system. In this paper we investigate the effects of a \emph{time delay} in the 
feedback for a protocol that performs an instantaneous maximization of the 
center-of-mass velocity. For the \emph{one} and the \emph{few particle} cases
the flux decreases with increasing delay, as an effect of the decorrelation of
the present state of the system with the information that the controller uses, 
but the delayed closed-loop protocol succeeds to perform better than its 
open-loop counterpart provided the delays are smaller than the 
characteristic times of the Brownian ratchet.
For the \emph{many particle} case, we also show that for small delays the 
center-of-mass velocity decreases for increasing delays. However, for 
large delays we find the surprising result that the presence of the delay 
can improve the performance of the nondelayed feedback ratchet and the flux
can attain the maximum value obtained with the optimal periodic protocol.
This phenomenon is the result of the emergence of a dynamical regime 
where the presence of the delayed feedback stabilizes one quasiperiodic
solution or several (multistability), which resemble the solutions 
obtained in the so-called threshold protocol.
Our analytical and numerical results point towards the feasibility of an
\emph{experimental implementation} of a feedback controlled ratchet that
performs equal or better than its optimal open-loop version.
\end{abstract}


\maketitle

\section{Introduction}
The ratchet effect consists of the emergence of a directed
transport in a spatially periodic system out of equilibrium
through the introduction of an external perturbation. The celebrated
ideas of rectifying thermal noise, originally introduced by
Smoluchowski~\cite{smo12} and later resumed by
Feynman~\cite{fey63}, were explicitly used in the context of
directed transport in the 1990s~\cite{ajd93,mag93,ast94}. Since
then, these systems have been studied due to its importance from a
theoretical point of view in nonequilibrium physics~\cite{rei02}
and its applications to many other fields such as
condensed matter or biology~\cite{rei02,lin02}.
\par

One of the main ratchet types are the flashing ratchets that
operate switching on and off a spatially periodic asymmetric
potential. A simple periodic or random switching is able to
achieve a rectification of thermal fluctuations and produce a net
current of particles. Recently, a new class of control protocols
that use instant information about the state of the system to take
the decision of switching on or off have been
introduced~\cite{cao04}. These so-called closed-loop or feedback
control protocols have been proven to be an effective way to
increase the net current in collective Brownian
ratchets~\cite{cao04,din05,fei06}. Feedback control can be
implemented in systems where particles are
monitored~\cite{rou94,mar02}. This monitoring gives information
about the position of the particles that can be used to switch on
or off the potential in real time according to a given protocol.
For instance, in Ref.~\cite{rou94} the motion of colloidal
particles induced by a sawtooth dielectric potential, which is
turned on and off periodically, is experimentally studied
monitoring the particles. This suggest that a feedback controlled
version of the ratchet in~\cite{rou94} can be constructed
gathering information about the state of the system with a charge coupled
device (CCD)
camera and using this information to decide whether to turn on and
off the potential in real time. In addition, feedback ratchets
have been recently suggested as a mechanism to explain the
stepping motion of the two-headed kinesin~\cite{bie07}.
\par

All Brownian feedback ratchets considered until now use instant
information to operate, that is, they all measure the state of the
system and act \emph{instantaneously} according to that
measurement. However, in realistic devices there is always a time
delay between the input measurements and the output control action
due to physical limitations to the velocity of transmission and
processing of the information~\cite{ste94,bec05}. For example, in
the construction of the feedback controlled version of the ratchet
in~\cite{rou94} time delays in the feedback will be present due to
the finite time needed to take a picture with a CCD camera,
transmit it, process it, and implement the resulting decision of
switching on or off the potential. Therefore it is important to
compute the effects of time delays in the feedback, because
it clarifies in which real ratchet systems it is
experimentally feasible to obtain the increase of velocity
predicted in~\cite{cao04}. The study of time-delayed feedback is
also relevant because it appears naturally in many stochastic
processes, such as complex systems with self-regulating
mechanisms~\cite{boc00,fra05b}. For another type of ratchets,
deterministic feedback ratchets, some of the effects of time delay
in the feedback have been studied \cite{kos05,son06}.
\par

In the current paper we investigate how a time delay in the
control of a feedback flashing ratchet affects the net flux. In
the next section we describe the ratchet model with the
time-delayed feedback control policy. In Sec.~\ref{sec:one} we
study in detail the case of one particle, getting an effective
potential description for the flux in the relevant case of small
time delays. We also present an alternative approach to
understanding the dependence of the flux with time delay in terms of
the covariance, and we describe the behavior for large delays. In
Sec.~\ref{sec:few} we treat the collective ratchet with few
particles and relate its center-of-mass velocity with the one
particle flux previously studied.
In Sec.~\ref{sec:many} we study the many particle ratchet, which exhibits
a somehow counterintuitive behavior; first we briefly review
the results for zero delays that will be useful, and thereafter we
expose the results in the two dynamical regimes of small delays and
large delays. Finally, all the results are summarized and
discussed in Sec.~\ref{sec:con}.
\par

\section{Model}\label{sec:model}

We consider $N$  overdamped Brownian particles at temperature $T$ in a ratchet
potential $V(x)$. The force acting on the \emph{i}th particle at position $x_i(t)$ is
$\alpha(t)F(x_i(t))$, where $F(x)=-V'(x)$ and $\alpha(t)$ implements the
action of the controller. Therefore the system dynamics is defined by the
Langevin equations
\begin{equation}\label{langevin}
\gamma \dot x_i(t)=\alpha(t)F(x_i(t))+\xi_i(t);\quad i=1,\dots,N,
\end{equation}
where $\gamma$ is the friction coefficient (related to the diffusion
coefficient $D$ through Einstein's relation $D=k_BT/\gamma$) and
$\xi_i(t)$ are Gaussian white noises of zero mean and variance $\langle
\xi_i(t)\xi_j(t^\prime)\rangle =2\gamma k_B T\delta_{ij}\delta(t-t^\prime)$.
\par

In order to study the effects of time-delayed feedback controls let us include
a time delay of value $\tau$ in the control of the paradigmatic
\emph{maximization of the center-of-mass instant velocity}
protocol~\cite{cao04}. The controller measures the sign of the net force per
particle
\begin{equation}
f(t)=\frac{1}{N}\sum_{i=1}^N F(x_i(t)),
\end{equation}
and, after a time $\tau$, it switches on the potential ($\alpha=1$)
if the net force was positive or it switches off ($\alpha=0$) if it was
negative. Thus the control protocol reads
\begin{equation}\label{alfa-delay}
  \alpha(t)=
  \begin{cases}
    \Theta(f(t-\tau)) &\text{if } t\geq \tau ,\\
    0 &\text{otherwise},
  \end{cases}
\end{equation}
with $\Theta$ the Heaviside function [$\Theta (x)=1$ if $x>0$, else $\Theta
(x)=0$].
\par

\begin{figure}
  \begin{center}
    \includegraphics [scale=0.5] {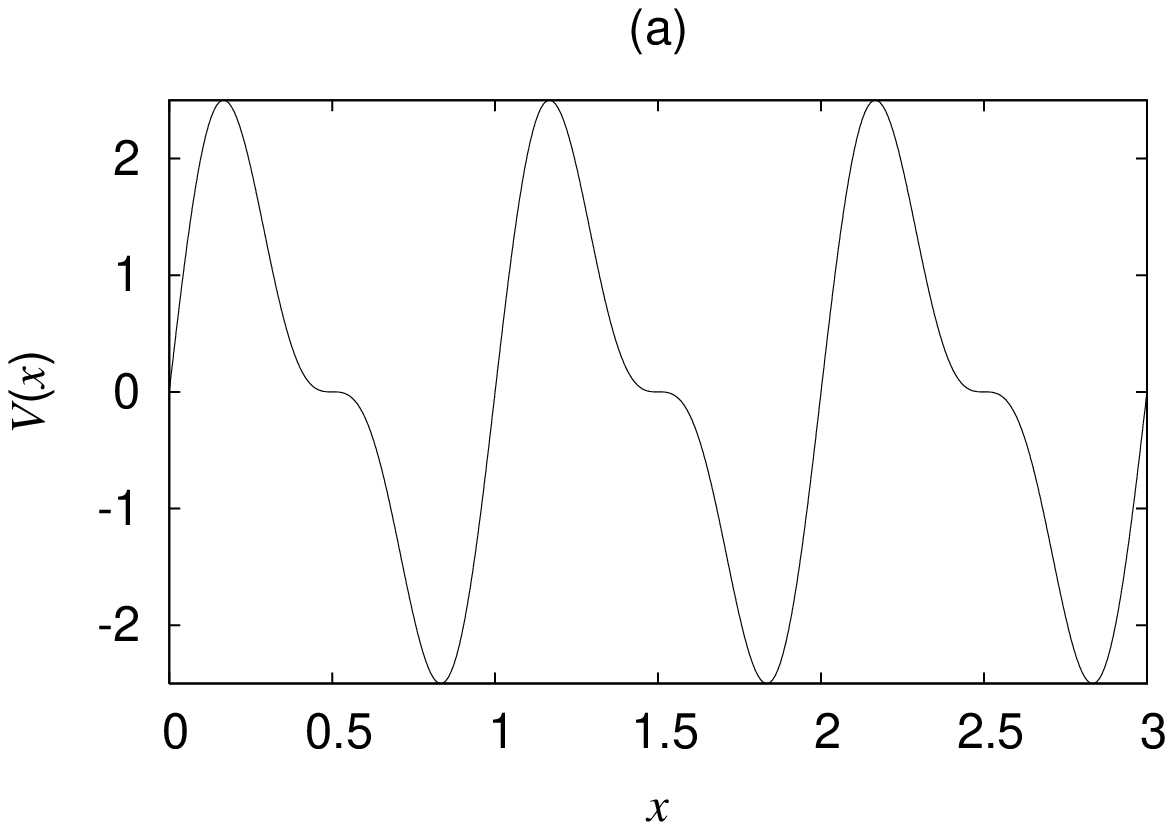}
    \includegraphics [scale=0.5] {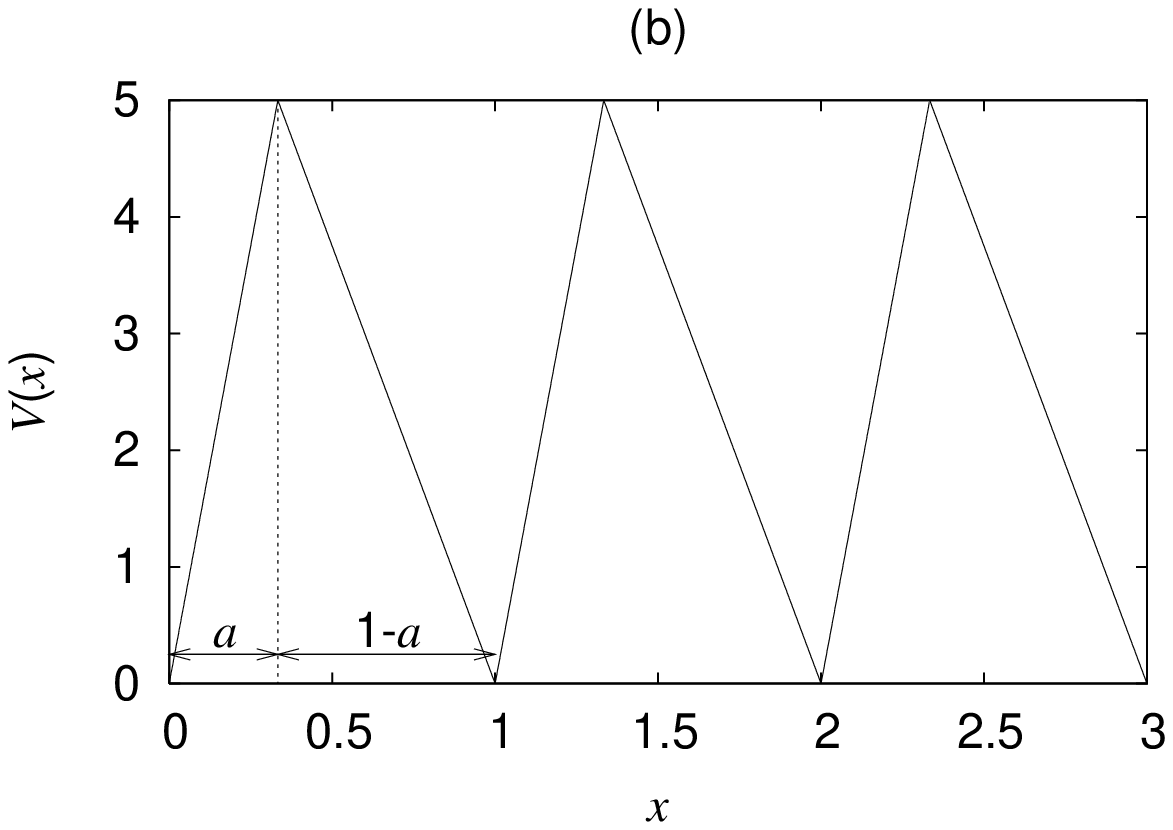}
  \end{center}
  \caption{Panel (a): `Smooth' potential \eqref{smoothpot} for $V_0=5k_BT$.
    Panel (b): `Sawtooth' potential \eqref{sawtoothpot} for $V_0=5k_BT$ and $
  a = 1/3$. Units: $L=1$, $k_BT=1$.
  }
  \label{fig:pot}
\end{figure}

Finally, to completely fix the model we choose a piecewise linear
sawtooth potential $V(x)=V(x+L)$ of height $V_0$ and asymmetry
parameter $a<1/2$,
\begin{equation} \label{sawtoothpot}
  V(x)=
  \begin{cases}
    \frac{V_0}{a}\frac{x}{L} &\text{if } 0\leq \frac{x}{L}\leq a ,\\
    V_0-\frac{V_0}{1-a}\left(\frac{x}{L}-a\right) &\text{if } a<
    \frac{x}{L}\leq 1.
  \end{cases}
\end{equation}
We have verified that the results found in this paper are valid
for other potentials provided they have the same height of the
potential $V_0$ and the same asymmetry parameter $a$, with $V_0$
defined as the difference between the maximum and the minimum
values of the potential and $aL$ as the distance between the
minimum and the maximum positions. For this verification we have
considered the `smooth' potential
\be \label{smoothpot}
V(x) = \frac{2V_0}{3\sqrt{3}} \left[ \sin \left(\frac{2\pi x}{L}
\right) + \frac{1}{2} \sin \left( \frac{4\pi x}{L} \right)
\right],
\ee
which has potential height $V_0$, period $L$, and asymmetry
$a=1/3$. See Fig.~\ref{fig:pot}.

In the study of these feedback ratchets it proves to be useful to
distinguish three cases: one particle, few particles, and many
particles. This classification is based on the results of the
zero delay studies \cite{cao04,din05,fei06}, which revealed
different characteristics and analytical approximations for each
case. The many particle case is formed by those feedback collective ratchets
that for zero delay have net force fluctuations smaller than the maximum
absolute value of the net force; see Refs.~\cite{cao04,din05,fei06}.
\par

Throughout the rest of this paper, we will use units where $L=1$,
$k_BT=1$, and $D=1$.

\section{One particle}\label{sec:one}

In this section we discuss the simpler case of a
ratchet consisting of one particle, so that the position $x(t)$ is
governed by Eq.~\eqref{langevin} with $N=1$ and $ \alpha(t) $ given by
Eq.~\eqref{alfa-delay}, which is a nonlinear stochastic delay differential
equation. In general, there is no analytical treatment for these
time-delayed stochastic equations. Here, we shall write the corresponding
delay Fokker-Planck equation \cite{gui99}, and use a perturbative technique
\cite{fra05,fra05b} to obtain an effective potential description for small
delays that leads to approximate analytical expressions for the flux. Finally,
in this section we shall get insight in the regime of large
delays by studying the covariance of the sign of the net force.
\par

\begin{figure}
  \begin{center}
    \includegraphics [scale=0.6] {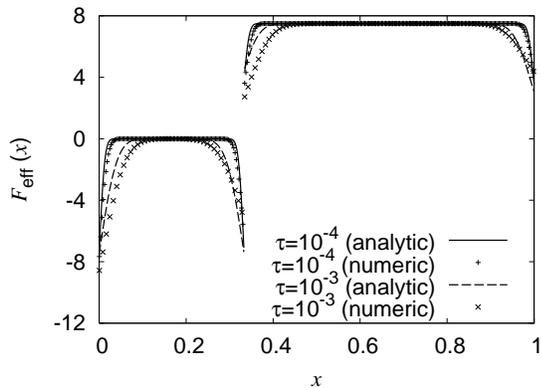}
  \end{center}
  \caption{Effective force for small delays for potential height
    $V_0=5k_BT$ and asymmetry $a=1/3$ in the one particle case
  [Eq.~\eqref{heff}]. Units: $L=1$, $D=1$, $k_BT=1$. }
  \label{fig:h_eff}
\end{figure}

The force that the particle feels with the
inclusion of the time
delay $\tau$ in the control [Eq.~\eqref{alfa-delay}] depends both on the actual
position $x:= x(t)$ and on the delayed position $x_\tau:= x(t-\tau)$. This
force $F_{\tau}(x,x_\tau)$ is periodic in both arguments,
$F_{\tau}(x,x_\tau)= F_{\tau}(x+1,x_\tau)=
F_{\tau}(x,x_\tau+1)$, and reads
\begin{equation}\label{Ftau}
  F_{\tau}(x,x_\tau)=
  \begin{cases}
    0&\text{if } 0\leq x_\tau\leq a,\\
    \frac{-V_0}{a} &\text{if } a< x_\tau\leq 1\text{ and } 0\leq x\leq a,\\
    \frac{V_0}{1-a} &\text{if } a< x_\tau\leq 1\text{ and } a<x\leq 1.\\
  \end{cases}
\end{equation}
In particular, $F_{\tau}(x,x)=:F_0(x)$
corresponds to the effective force of the instant maximization control
protocol without delay~\cite{cao04}, i.e.,
\begin{equation}
  F_{0}(x)=
  \begin{cases}
    0 &\text{if } 0\leq x\leq a ,\\
    \frac{V_0}{1-a} &\text{if } a<x\leq 1.\\
  \end{cases}
\end{equation}
In terms of the force~\eqref{Ftau}, the evolution of the position
of the particle obeys the stochastic delay differential equation
\begin{equation}
\dot x(t)=F_{\tau}(x(t),x(t-\tau))+\xi(t).
\end{equation}
The probability density $\rho(x,t)$ of this stochastic process satisfies
a delay Fokker-Planck equation~\cite{gui99,fra02,fra05b,fra05}, which involves
the two-point probability density as follows:
\begin{equation}
\begin{split}
\frac{\partial}{\partial x}\rho(x,t) & =
-\frac{\partial}{\partial x}
\int F_{\tau}(x,x_\tau)\rho(x,t;x_\tau,t-\tau)\;dx_\tau\\
& \quad  +\frac{\partial^2}{\partial x^2}\rho(x,t).
\end{split}
\end{equation}
For small delays, this equation can be treated perturbatively; then, following
Refs.~\cite{fra05,fra05b}, the explicit effective force for small delays can
be achieved by computing
\begin{equation}\label{integ}
  \Feff(x)=\int F_\tau(x,x_\tau)P(x_\tau,t+\tau|x,t)\; dx_\tau,
\end{equation}
where the short time propagator $P(x,t+\tau|x,t)$ (see \S 4.4.1 in
Ref.~\cite{ris89}) is
\begin{equation}
  P(x_\tau,t+\tau|x,t)=\frac{1}{\sqrt{2\pi\tau}}\exp\left(
-\frac{[x_\tau-x-F_0(x)\tau]^2}{2\tau}
\right).
\end{equation}
Due to the Gaussian form of the propagator in this small delay approximation,
we can neglect the long tails of the Gaussian propagator and restrict the
integration in Eq.~\eqref{integ} to the intervals $(a-1,1)$ and $(0,1+a)$ for
$0\leq x\leq a$ and $a<x\leq 1$, respectively. We get
\begin{widetext}
\begin{equation}\label{heff}
\Feff(x)=\Feff(x+1)=
\begin{cases}
  -\frac{V_0}{2a}\left[\erfc\left(\frac{x}{\sqrt{2\tau}}\right)+
    \erfc\left(\frac{a-x}{\sqrt{2\tau}}\right)\right] &\text{if } 0\leq
  x\leq a,\\

  \frac{V_0}{2(1-a)}\left[2-\erfc\left(\frac{1-x-\frac{V_0\tau}{1-a}}{\sqrt{2\tau}}\right)-\erfc\left(\frac{x-a+\frac{V_0\tau}{1-a}}{\sqrt{2\tau}}\right)\right]
  &\text{if } a<
  x\leq 1,\\
  \end{cases}
\end{equation}
\end{widetext}
where $\erfc(x)$ is the complementary error function. On the other
hand, the value of the effective force can be computed numerically
by splitting in bins the position of the particle
and evaluating the probability of being in those bins. For small
delays, Eq.~\eqref{heff} gives a good estimation as shown in
Fig.~\ref{fig:h_eff}.
\par

The main effect of the inclusion of a small delay in the control
is a slant of the effective force near the points of
discontinuity. This effect lies on the idea that the closer the particle is to
the discontinuities, the more
probable is that the controller makes a mistake. For instance,
when the particle is to the left of $x=a$ and close to it, there
are two possibilities: (i) if the retarded position was to the
left too then the controller sets the potential off, and (ii) if
the retarded position was to the right then the controller sets
the potential on and the particle feels a negative force $-V_0/a$.
Therefore in the points to the left of $x=a$ and close to it the
force takes an effective value between $0$ and $-V_0/a$, resulting
in a negative effective force.
\par

\begin{figure}
\begin{center}
\includegraphics [scale=0.6] {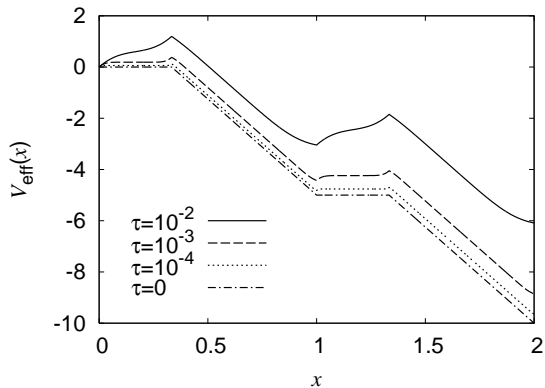}
\end{center}
\caption{Effective potential [Eq.~\eqref{pot-sol}] for potential height
  $V_0=5k_BT$ and  asymmetry $a=1/3$ for time delays
 $\tau=0$, $10^{-4}$, $10^{-3}$, and  $10^{-2}$.
 Units: $L=1$, $D=1$, $k_BT=1$.
 }
  \label{veff}
\end{figure}

\begin{figure}
\begin{center}
\includegraphics [scale=0.6] {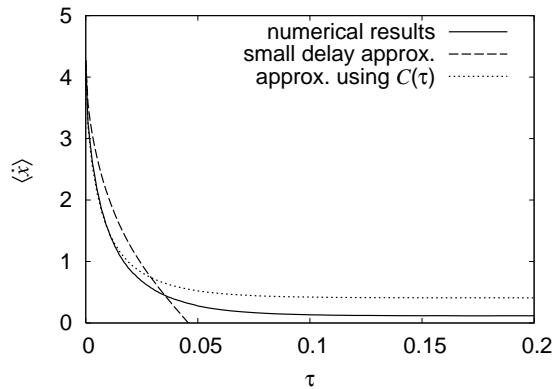}
\end{center}
\caption{One particle flux vs delay. The numerical
  result (solid line) is compared with the estimations using
  the effective potential~\eqref{pot-sol} (dashed line) and using $C(\tau)$
  (dotted line). $V_0=5k_BT$ and $a=1/3$. Units: $L=1$, $D=1$, $k_BT=1$.}
  \label{fig:flux_vs_delay}
\end{figure}

In this effective description the position of the particle evolves
with a Langevin equation $\dot x(t)=\Feff(x)+\xi(t)$, with the
associated (nondelayed) effective Fokker-Planck equation
\begin{equation}
  \frac{\partial}{\partial t}\rho(x,t)=-
  \frac{\partial}{\partial x}\left[\rho(x,t)\Feff(x)\right]+
  \frac{\partial^2}{\partial x ^2}\rho(x,t),
\end{equation}
with periodic boundary conditions. The average velocity is obtained computing
the expectation value of the velocity in the stationary distribution of the
effective Fokker-Planck equation~\cite{rei02}:
\begin{equation}\label{FPE-sol}
  \vone=\frac{1-e^{\Veff(1)-\Veff(0)}}{\int_0^1 dx \int_x^{x+1} dy
  \;e^{\Veff(y)-\Veff(x)}},
\end{equation}
where $\Veff(x)=-\int_0^x \Feff(s)\;ds$. Integrating
$\Feff(x)$ we get the expression of the approximate effective
potential for small delays $\Veff(x)$,
\begin{widetext}
\begin{equation}\label{pot-sol}
\Veff(x)=
\begin{cases}
  \frac{V_0\sqrt{2\tau}}{2a}\Big[
    \ierfc\left(\frac{a}{\sqrt{2\tau}}\right)
  +\ierfc\left(\frac{x}{\sqrt{2\tau}}\right)
  -\ierfc\left(\frac{a-x}{\sqrt{2\tau}}\right)
  -\frac{1}{\sqrt{\pi}}
\Big]
&\text{if } 0\leq
x\leq a,
\vspace{2mm}
\\
\frac{V_0\sqrt{2\tau}}{2(1-a)}\Bigg[
  \frac{2(a-x)}{\sqrt{2\tau}}
  +\ierfc\left(\frac{1-x-\frac{V_0\tau}{1-a}}{\sqrt{2\tau}}\right)
  -\ierfc\left(\frac{1-a-\frac{V_0\tau}{1-a}}{\sqrt{2\tau}}\right)
  +\ierfc\left(\frac{\frac{V_0\tau}{1-a}}{\sqrt{2\tau}}\right)\\
  \hspace{3cm}
  -\ierfc\left(\frac{x-a+\frac{V_0\tau}{1-a}}{\sqrt{2\tau}}\right)
  +\frac{2(1-a)}{a}\ierfc\left(\frac{a}{\sqrt{2\tau}}\right)
  -\frac{2(1-a)}{a\sqrt{\pi}}
\Bigg]
&\text{if } a<
x\leq 1,\\
  \end{cases}
\end{equation}
\end{widetext}
in the interval $[0,1]$, and outside
$V_{\text{eff}}(x)=V_{\text{eff}}(y)+(x-y)V_{\text{eff}}(1) $, with
$y\equiv x\mod 1$, $y\in[0,1]$. The function $\ierfc$ is the
first iterated integral of the complementary error
function~\cite{abra},
\begin{equation}
\ierfc(x)=\int_x^\infty \erfc(s)\;ds=-x\;\erfc(x)+\frac{e^{-x^2}}{\sqrt{\pi}}.
\end{equation}
This effective potential is depicted in Fig.~\ref{veff}, where we see that an
increase of the delay implies a decrease of the average tilt of the potential.
Eventually, the stationary flux is calculated inserting
Eq.~\eqref{pot-sol} in Eq.~\eqref{FPE-sol}. The resulting
approximate expression gives good results for very small delays
and a good estimate of the decrease rate for small delays. See
Fig.~\ref{fig:flux_vs_delay}. [This can be understood noting that
although for some positions the corrections to the effective force
are appreciable already for quite small delays (see
Fig.~\ref{fig:h_eff}) this only happens in small space intervals
and therefore the results for the flux are better than expected.]
The approximate analytical expression obtained gives the average
velocity in terms of the main magnitudes of the system, namely,
the height of the potential $V_0$, its asymmetry $a$, and the
time delay in the feedback $\tau$. We have checked that this
result is in good agreement also for other potentials.

\par

\begin{figure}
\begin{center}
\includegraphics [scale=0.6] {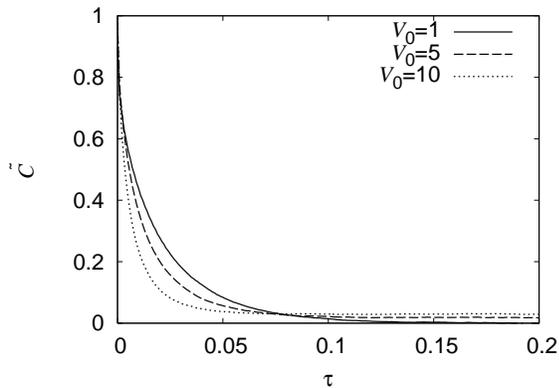}
\end{center}
\caption{Covariance $\widetilde C$
  [Eq.~\eqref{correlcentered}] as a function of the time delay for potential
  heights $V_0=k_BT$, $5k_BT$, and $10k_BT$. Asymmetry
  parameter $a=1/3$. Units: $L=1$, $D=1$, $k_BT=1$.
}
  \label{fig:cor-1}
\end{figure}

Another approach can be taken to understand the observed decrease in the
flux for increasing delay. The instant maximization protocol does
not use detailed  information about the position of the particles, it simply
deals with the sign of the net force, namely, $\sgn f$, [with $ \sgn(x) =
1 $ for $x>0$, $ \sgn(x) = 0 $ for $x=0$, and $ \sgn(x) = -1 $ for
$x<0$]. The flux performance of the protocol would be optimal if it
would have received the present sign of the net force, $\sgn f(t)$, but it
does receive its value a time $\tau$ earlier, $\sgn f(t-\tau)$.
This earlier value contains information about the present
value because both values are correlated, as can be shown
computing the covariance
\begin{equation}\label{correlcentered}
\begin{split}
\widetilde{C}(\tau):=& \Bigl\langle [\sgn f(t)-\mu][\sgn
f(t-\tau)-\mu]\Bigr\rangle \\
=& C(\tau) - \mu^2,
\end{split}
\end{equation}
where
\begin{equation}
\begin{split}
C(\tau) :=& \langle \sgn f(t) \sgn f(t-\tau)\rangle, \\
\mu :=& \langle\sgn f(t)\rangle = \langle\sgn f(t-\tau)\rangle.
\end{split}
\end{equation}
The decrease of the function $\widetilde{C}(\tau)$ for
increasing $\tau$ (Fig.~\ref{fig:cor-1}) explains the
decrease of the center-of-mass velocity as a consequence of the loss of
information about the present sign of the net force. In addition, we can
obtain an estimation of the flux decrease with the following heuristic
argument. Let us calculate the covariance using
\begin{equation}\label{corrmu}
\begin{split}
 {C} & =P_{++}+P_{--}-P_{+-}-P_{-+},\\
 \mu & = P_{++}-P_{--}+P_{+-}-P_{-+},
\end{split}
\end{equation}
where $P_{ij}$ is the joint probability of having a positive
($i=+$) or negative ($i=-$) net force at time $t$ and a positive
($j=+$) or negative ($j=-$) net force at time $t-\tau$. These
probabilities can be computed if we assume that the system
performance can be explained with the simplified description that
$\sgn f(t-\tau)$ is different from $\sgn f(t)$ with probability $p$.
This description allows us to use the results found in
\cite{cao07} for the instantaneous maximization
protocol with a controller receiving $\sgn f(t)$ through a noisy
channel with noise level $p$. In fact,
notice that the plot of the effective potential (Fig.~\ref{veff})
resembles the form of the effective potential found in Ref.~\cite{cao07} for
the noisy channel. This elementary description gives the values $P_{-+}= bp$,
$P_{--}=b(1-p)$, $P_{+-}=(1-b)p$ and $P_{++}=(1-b)(1-p)$ for the joint
probabilities, with $b$ the probability of $\sgn f(t)$ being negative. Thus
Eqs.~\eqref{corrmu} can be rewritten in terms of the probability of error
$p=P_{+-}+P_{-+}$ and the probability $b=P_{-+}+P_{--}$ as
\begin{equation}\label{corrmuval}
\begin{split}
 {C} & \approx 1-2p,\\
 \mu & \approx 1-2b.
\end{split}
\end{equation}
In~\cite{cao07} it is shown that for small potential heights
(small $V_0$)  $ b \approx a $ and
\begin{equation}
  \vone \approx V_0 (1-2p).
\end{equation}
Therefore this simplified description suggests
\begin{equation} \label{voneC}
  \vone \sim V_0 C.
\end{equation}
For larger potential heights, a better estimation is obtained
evaluating the general expression $\vone (p)$ of Ref.~\cite{cao07}
at $p(\tau)=[1-C(\tau)]/2$. This estimation is plotted in
Fig.~\ref{fig:flux_vs_delay}, where it is compared with numerical
results and the analytical small delay approximation
[Eqs.~\eqref{FPE-sol} and~\eqref{pot-sol}].
\par

\begin{figure}
\begin{center}
\includegraphics [scale=0.6] {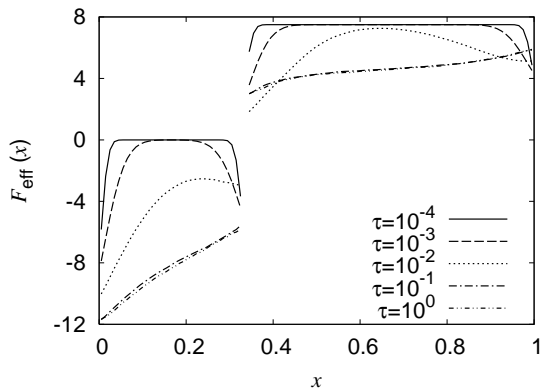}
\end{center}
\caption{Effective force (from numerical simulations)
  ($N=1$) for several time delays. $V_0=5k_BT$ and $a=1/3$. Units: $L=1$,
  $D=1$, $k_BT=1$.}  \label{fig:Feff_large_delays}
\end{figure}

The average velocity of the particle for large delays is not zero,
but reaches a constant value independent of the delay (see
Fig.~\ref{fig:flux_vs_delay}). We have seen that the function
$C(\tau)$ also tends to a constant nonzero value in the same
characteristic time that the velocity does, in qualitative
agreement with the estimation described after Eq.~\eqref{voneC},
although this estimation does not give
the correct value of the flux. Therefore this estimation gives
good quantitative results for small delays and the qualitative
behavior for large delays. The large $\tau$ behavior observed for
the flux implies an effective force independent of the time delay
$\tau$ for large enough values of $\tau$, as we show in
Fig.~\ref{fig:Feff_large_delays}. The average over $x$ of the
numerical large $\tau$ effective force is positive, in agreement
with the positive net flux obtained. For example, for asymmetry
parameter $a=1/3$ and potential heights $V_0=1$, $ 5 $, and $ 10
$, the net flux is $\vone_{\tau\to\infty}\approx 0.01$, $0.12$, and
$0.18$, respectively. The convergence to this constant
value can be understood realizing that the covariance $\widetilde
C$ becomes negligible for large delays, i.e., the fluctuations of
$\sgn f$ around its mean value at $t$ and at $t-\tau$ are
independent. This indicates that the system dynamics is
effectively the same as that for an open-loop control protocol, as
the correlation between the switches and the state of the system
are negligible.
\par

Comparing the results for the delayed instant maximization
protocol with the optimal periodic open-loop protocol, we see that the former
performs better than the latter even for nonzero delay, provided the delay
is smaller than the characteristic times of the dynamics of the Brownian
ratchet. Therefore the instant maximization protocol gives a larger flux
than the optimal open-loop control protocol for time delays $\tau$ such that
$\tau\ll {\cal T}_{\text{on}}$, $\tau\ll {\cal T}_{\text{off}}$, where ${\cal
  T}_{\text{on}} \sim {(1-a)^2}/{V_0}$ and ${\cal T}_{\text{off}} \sim
{a^2}/{2}$ are the on-potential and off-potential times in the optimal periodic
protocol~\cite{ast94}. See, for example, Fig.~\ref{fig:flux_vs_delay}  and
compare with $ \vone_{\text{open}} \approx 0.3$ that is the value for the
optimal periodic protocol for $V_0=5$ and $a=1/3$, which has ${\cal
  T}_{\text{on}}\approx 0.06$ and ${\cal T}_{\text{off}}\approx 0.05$.

\section{Few particles}\label{sec:few}

In this section we deal with a collective ratchet compounded of a
few particles. We will show that the center-of-mass velocity for the few
particle case can be related with the velocity obtained for the one particle
ratchet studied in the section before.
\par

\begin{figure}
\begin{center}
\includegraphics [scale=0.6] {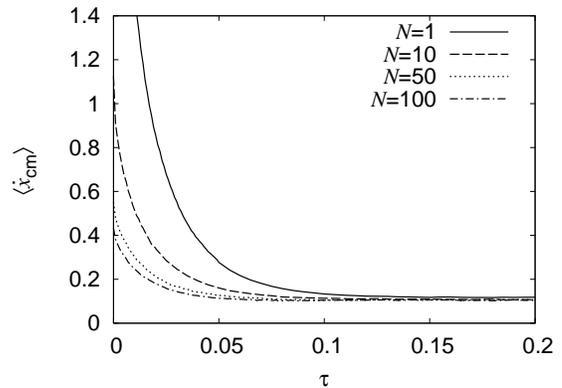}
\end{center}
\caption{Center-of-mass velocity as a function of the time delay for number of
  particles $N=1$, $10$, $50$, and $100$. Parameters of the potential:
  $V_0=5k_BT$ and $a=1/3$. Units: $L=1$, $D=1$, $k_BT=1$.}
  \label{fig:vcm_few}
\end{figure}

As in the one particle case, the effect of the inclusion of a
delay is a decrease in the covariance  and in the center-of-mass velocity
(Fig.~\ref{fig:vcm_few}). Therefore we can also interpret the
decrease in the center-of-mass velocity as a consequence of the
loss of information of the present sign of the net force, and then
assume that the system performance can be explained with the
simplified description that $\sgn f(t-\tau)$ is different from $\sgn
f(t)$ with probability $p$. This simplified description leads for small
potential heights~\cite{cao07,fei07} to
\begin{equation}\label{corrmuvalN}
\begin{split}
 {C}_N(\tau) & \sim 1-2p_N,\\
 \mu_N & \sim 1-2b_N,\\
 \vmedia_N &\approx\frac{V_0(1-2p_N)}{\sqrt{2\pi a(1-a)N}}\sim
  \frac{V_0 C_N}{\sqrt{2\pi a(1-a)N}},
\end{split}
\end{equation}
where the subscript $N$ denotes that the quantities are the values
in the case of $N$ particles. We have numerically found that the function
$C_N(\tau)$ is approximately the same for any number of particles in this
regime of a few particles, and $C_N \sim C$. Thus we have the relation
\begin{equation}\label{v-few}
  \vmedia_N(\tau)\approx\frac{\vone(\tau)}{\sqrt{2\pi a(1-a)N}}
\end{equation}
between the velocities for one and for $N$ particles for a given
delay $\tau$. This Eq.~\eqref{v-few} gives good results for
small values of the delay. In particular, inserting
Eq.~\eqref{FPE-sol} in Eq.~\eqref{v-few} we obtain an analytical approximate
expression for the case of few particles in the regime of small delays.
\par

We stress that,
analogously to the zero delay case~\cite{cao04}, the main effect
of having a collective ratchet is a decrease in the magnitude  of
the force fluctuations. This fact gives a center-of-mass velocity
inversely proportional to the square-root of the number of
particles, as Eq.~\eqref{v-few} states. Thereby, if the number $N$ of particles
increases, there will be a decrease of the values of the delay that give
better performances for the delayed instant maximization protocol than for the
optimal periodic protocol.
\par

On the other hand, for large time delays the analogy between the delayed
protocol and the noisy channel protocol no longer gives a good estimate. In
this regime of large delays the covariance $\widetilde C(\tau) = C(\tau)-\mu^2
$ becomes negligible indicating that $ \sgn f(t-\tau) $ and $ \sgn f(t) $ are
nearly uncorrelated and that the system effectively behaves as if it were
driven by an effective open-loop control protocol.
In addition, we observe that the value of the
center-of-mass velocity becomes independent of the
number of particles (see Fig.~\ref{fig:vcm_few}). This is a
hallmark of collective open-loop control ratchets, in which the absence
of feedback decouples the Langevin equations provided the particles do
not explicitly interact with each other.

\section{Many particles}\label{sec:many}

We study here the effects of time delays in the feedback
controlled Brownian ratchet described in Sec.~\ref{sec:model} for
the many particle case, considering both the `smooth' potential
and the `sawtooth' potential for various potential heights and
different initial conditions.

\begin{figure}
  \begin{center}
    \includegraphics [scale=0.6] {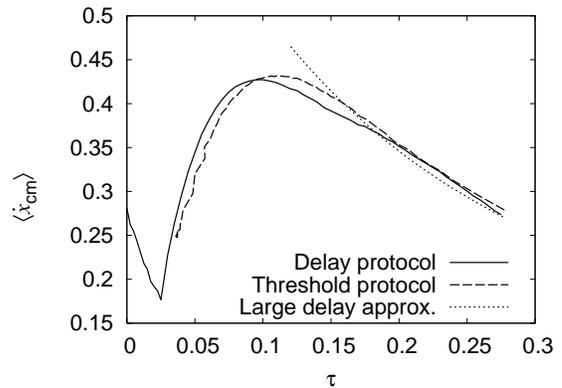}
  \end{center}
  \caption{Many particle case: Center-of-mass velocity as a function of the delay (for large
  delays only the first branch is represented here), and comparison with the
  results obtained with the threshold protocol and with the large delay
  approximation Eq.~\eqref{vlargetau}. For the `smooth' potential
  \eqref{smoothpot} with $V_0=5k_BT$ and $ N = 10^5 $ particles.
  Units: $L=1$, $D=1$, $k_BT=1$.
  }
  \label{fig:taumin}
\end{figure}

We find that the system presents two regimes separated by a delay
$\taumin$ for which the center-of-mass velocity has a minimum; see
Fig.~\ref{fig:taumin}. In the small delay regime ($ \tau < \taumin
$) the flux decreases with increasing delays as one could expect.
On the contrary, in the large delay regime ($ \tau > \taumin $) we
have observed and explained a surprising effect, namely, the
center-of-mass velocity increases for increasing delays and the
system presents several stable solutions. We have found that this
critical time delay $ \taumin $ is inversely proportional to the
potential height $ \taumin \propto 1/V_0 $ with a proportionality
constant that mildly depends on the number of particles.

\subsection{Zero delay}\label{sec:zero}

The many particle ratchet in absence of delay (i.e., $\tau=0$ in
the model of Sec.~\ref{sec:model}) has been studied in
Ref.~\cite{cao04}. It has been shown that the net force per
particle exhibits a quasideterministic behavior that alternates
large periods of time $t_{\text{on}}$ with $ f(t)>0 $ (on
dynamics) and large periods of time $t_{\text{off}}$ with $ f(t)<0
$ (off dynamics). The center-of-mass velocity can be computed as
\be \label{vmediacero}
\vmedia = \frac{\Delta x(\ton)}{\ton+\toff},
\ee
with
\be \label{deltaxon}
\Delta x(\ton) = \Delta x_{\text {on}} [ 1-e^{-\ton/(2\Delta
t_{\text {on}})} ],
\ee
where $\Delta x_{\text{on}}$ and $ \Delta t_{\text{on}} $ are
obtained fitting the displacement during the `on' evolution for an
infinite number of particles (see Ref.~\cite{fei06} for details).

\begin{figure}
  \begin{center}
    \includegraphics [scale=0.6] {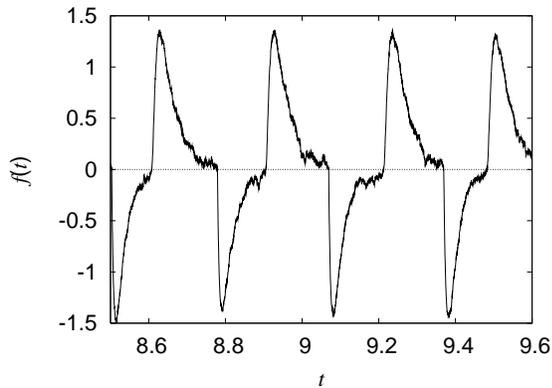}
  \end{center}
  \caption{Many particle case: Evolution of the net force
  with zero delay ($\tau=0$) for the `smooth' potential
  Eq.~\eqref{smoothpot} with  $V_0=5k_BT$ and
  $N=10^5$ particles. Units: $L=1$, $D=1$, $k_BT=1$.
  }
  \label{fig:evol0}
\end{figure}

On the other hand, for many particles the fluctuations of the net
force are smaller than the maximum value of the net force (see
Fig.~\ref{fig:evol0}). This allows the decomposition of the
dynamics as the dynamics for an infinite number of particles plus
the effects of the fluctuations due to the finite value of $N$.
The late time behavior of the net force $f(t)$ for an infinite
number of particles is given for the on and off dynamics by
\cite{cao04}
\be
f_\nu^\infty(t) = C_\nu e^{-\lambda_\nu(t-\tau_\nu)} \mbox{ with }
\nu = \mbox{on, off}.
\ee
The coefficients $ C_\nu $, $\lambda_\nu$, and $\tau_\nu$ can be
obtained fitting this expression with the results obtained
integrating a mean field Fokker-Planck equation obtained in the
limit $ N \to \infty $ and without delay; see
Refs.~\cite{cao04,fei06} for details. For a finite number of
particles the fluctuations in the force induce switches of the
potential and the times on and off are computed equating $
f^\infty_\nu $ to the amplitude of the force fluctuations,
resulting in \cite{cao04}
\be \label{tonoffzero}
\ton + \toff = b + d \ln N,
\ee
with $ b = \Con + \Coff $ and $ d = (\lambdaon + \lambdaoff) /
(2\lambdaon\lambdaoff) $.

\subsection{Small delays}\label{sec:small}

\begin{figure}
  \begin{center}
    \includegraphics [scale=0.6] {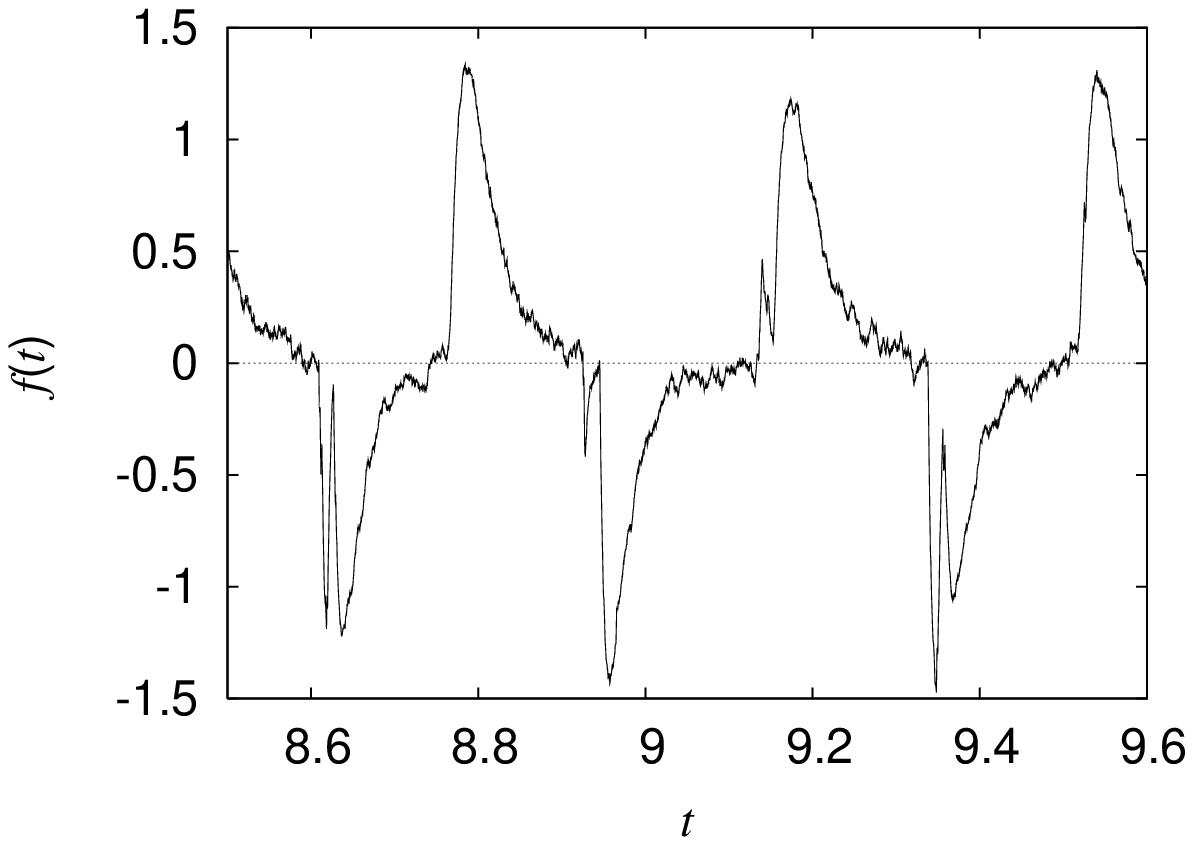}
  \end{center}
  \caption{Many particle case: Evolution of the net force with a small delay ($\tau=0.02$)
  for the `smooth' potential
  Eq.~\eqref{smoothpot} with  $V_0=5k_BT$ and
  $N=10^5$ particles. Units: $L=1$, $D=1$, $k_BT=1$.
  }
  \label{fig:evol}
\end{figure}

For small delays, $ \tau < \taumin $, we observe that the flux
decreases with the delay. See Fig.~\ref{fig:taumin}. We have seen
that this decrease is slower than that found for the few particle
case (Sec.~\ref{sec:few}), and that the expressions derived to
describe this decrease in the few particle case does not hold
here. However, the decrease observed here can be understood noting
that a change in the sign of $f(t)$ is perceived by the controller
a time $\tau$ after, what delays the reaction of the system and
makes the tails of $f(t)$ longer and implies an increase of the
time interval between switches. In addition, the form of $ f(t) $
is less smooth than for no delay because the delayed reaction of
the controller allows to have several sign flips in the $f(t)$
tails before the system reacts. This sign flips give short epochs
of fast switches of the potential (between long on and off
epochs), which lead to large fluctuations in $f(t)$. These large
fluctuations eventually destabilize these long period solutions
for $ \tau \sim \taumin $. See Fig.~\ref{fig:evol}.

As the main effect of the delay is to stretch the `on' and `off'
times of the dynamics, using the many particle approximation
\cite{cao04} we can write
\begin{equation}\label{small-del}
    \vmedia=\frac{\Delta x_{\text{on}}}{\ton+\toff+\Delta\tau}=
    \frac{\Delta x_{\text{on}}}{b+d \ln N +\Delta\tau},
\end{equation}
where we have found that the increase of the length of the on-off
cycle $ \Delta\tau $ is proportional to the delay $ \Delta\tau
\propto \tau$.

\subsection{Large delays}\label{sec:large}

\begin{figure}
  \begin{center}
    \includegraphics [scale=0.6] {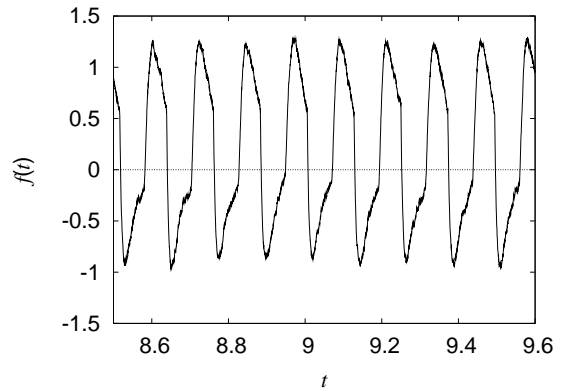}
  \end{center}
  \caption{Many particle case: Evolution of the net force with a large delay ($\tau=0.12$)
  for the `smooth' potential Eq.~\eqref{smoothpot} with  $V_0=5k_BT$ and
  $N=10^5$ particles. Units: $L=1$, $D=1$, $k_BT=1$.
  }
  \label{fig:evollarge}
\end{figure}

After the minimum flux is reached for $ \tau = \taumin $, the flux
begins to increase with the time delay (see
Fig.~\ref{fig:taumin}). This increase is due to a change in the
dynamical regime: for $ \tau > \taumin $ the present net force
starts to be nearly synchronized with the net force a time $ \tau
$ ago. This
\emph{self-synchronization} gives rise to a quasiperiodic solution of
period ${\cal T} = \tau $. Note that there is not a strict periodicity
due to stochastic fluctuations in the `on' and `off' times.
Looking at the $f(t)$ dependence, Fig.~\ref{fig:evollarge}, we see
that the solutions stabilized by the self-synchronization are
similar to those obtained with the threshold protocol
\cite{din05,fei06}. In Fig.~\ref{fig:taumin} we show that the
threshold protocol that has the same period gives similar
center-of-mass velocity values, confirming the picture.
(Differences are due to the fact that we have considered for the
threshold protocol simulations with on and off thresholds of the
same magnitude, while Fig.~\ref{fig:evollarge} shows that the
effective thresholds are different.)

This picture allows one to understand the increase of velocity for
increasing delay, and the presence of a maximum. This maximum is
related with the optimal values of the thresholds that have been
shown in \cite{fei06} to give a quasiperiodic solution of period
${\cal T}={\cal T}_{\text{on}} + {\cal T}_{\text{off}} $, with
${\cal T}_{\text{on}} $ and $ {\cal T}_{\text{off}} $ the optimal
`on' and `off' times of the periodic protocol. Therefore if we know
the values of $ {\cal T}_{\text{on}} $ and $ {\cal T}_{\text{off}}
$ for the optimal periodic protocol [${\cal T}_{\text{on}} \sim
(1-a)^2/V_0 $ and $ {\cal T}_{\text{off}} \sim a^2/2 $] we can
predict that the maximum of the center-of-mass velocity is reached
for a delay
\be
\tau_{\text{max}} = {\cal T}_{\text{on}} + {\cal T}_{\text{off}},
\ee
and has a value
\be
\vmedia_{\text{closed}} (\tau_{\text{max}}) =
\vmedia_{\text{open}}^{\text{max}},
\ee
with $ \vmedia_{\text{open}}^{\text{max}} $ the center-of-mass
velocity for the optimal open-loop protocol. Thus this expression
gives the position and height of the maximum of the delayed
feedback control protocol in terms of the characteristic values of
the optimal open-loop control. In particular, it implies that the
position and height of the maximum for the flux is independent of
the number of particles.

As an example we can apply these expressions to the `smooth'
potential with $V_0 = 5$ that for the optimal periodic protocol
gives $\vmedia = 0.44$ for ${\cal T}_{\text{on}}=0.06$ and ${\cal
T}_{\text{off}}=0.05$, so we obtain
$\tau_{\text{max}}=0.06+0.05=0.11$ in agreement with
Fig.~\ref{fig:taumin}.

\bigskip

\begin{figure}
  \begin{center}
    \includegraphics [scale=0.6] {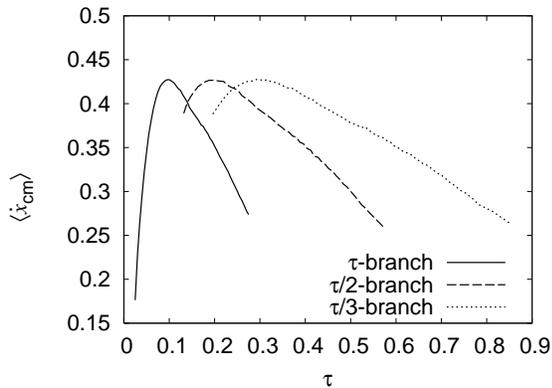}
  \end{center}
  \caption{Many particle case: First three branches of stable solutions
  for the `smooth' potential \eqref{smoothpot}
  with $V_0=5k_BT$ and $ N = 10^5 $ particles. Units: $L=1$, $D=1$, $k_BT=1$.
  }
  \label{fig:vbranches}
\end{figure}

\begin{figure}
  \begin{center}
    \includegraphics [scale=0.6] {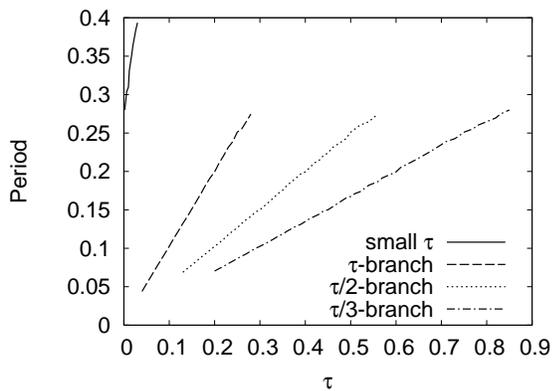}
  \end{center}
  \caption{Many particle case: Period ${\cal T}$ of the quasiperiodic solutions
    for small delays and for large delays. For large delays only the first 
    three branches of stable solutions are shown. `Smooth'
    potential~\eqref{smoothpot} 
    with $V_0=5k_BT$ and $ N=10^5$ particles. Units: $L=1$, $D=1$, $k_BT=1$.
  }
  \label{fig:Tbranches}
\end{figure}

For values of the delay of the order of or larger than $
\tau_{\text{max}} $ quasiperiodic solutions of other periods start
to be stable; see Fig.~\ref{fig:vbranches}. The periods for the
net force $f(t)$ that are found are those that fit an integer
number of periods inside a time interval $\tau$, verifying that
the present net force is synchronized with the net force a time
$\tau$ ago, that is, the quasiperiodic solutions have periods ${\cal T}
= \tau/2 $, $ {\cal T} = \tau/3 $, $\ldots$ In addition, it can be seen
that the center-of-mass velocity of the $n$ branch
$\vmedia_{\tau/n}$ whose $f(t)$ has period ${\cal T} = \tau/n $ is
related with that of the ${\cal T} = \tau $ branch through
\be \label{vmedian}
\vmedia_{\tau/n}(\tau) = \vmedia_\tau (\tau/n).
\ee
We highlight that several branches can be stable for the same time
delay $\tau$. Whether the system finally goes to one or another
stable solution depends on the initial conditions and on the
particular realization of the noise. See Figs.~\ref{fig:vbranches}
and \ref{fig:Tbranches}. For these branches we have found initial
conditions that go to these solutions and that remain in them
during several thousands of periods, indicating that they are
stable solutions or at least metastable solutions with a large
lifetime.

The analogy with the threshold protocol allows one to use the analytic
results of \cite{fei06} to get further insight in the numerical
results. The behavior for large delays for the ${\cal T}=\tau $ branch
can be obtained using the relation
\be \label{vlargetau}
\vmedia = \frac{\Delta x(\tau)}{\tau},
\ee
with $ \Delta x(\tau) $ given by Eq.~\eqref{deltaxon}. This
equation gives a good prediction for the largest delays of the
first branch (see Fig.~\ref{fig:taumin}).

On the other hand, for very large values of the delays of the
first branch the solutions in a given branch start to become
unstable, which can be understood noting that this happens when the
fluctuations of the net force become of the order of the absolute
value of the net force. Thus the maximum delay that gives a
stable solution in the first branch is
\be \label{tauinst}
\tau_{\text{inst}} = \ton+\toff = b+d \ln N,
\ee
where $b$ and $d$ are determined as in Eq.~\eqref{tonoffzero}. For
example, for the `smooth' potential with $ V_0 = 5 $, which has $
b = -0.070 $ and $d=0.031$, we obtain for $N = 10^5$ particles the
value $\tau_{\text{inst}}=0.29$ in accordance with the numerical
results shown in Figs.~\ref{fig:taumin} and \ref{fig:vbranches}.

The previous results for the first branch, Eqs.~\eqref{vlargetau}
and \eqref{tauinst}, can be extended to other branches by direct
application of the relation \eqref{vmedian}.

\section{Conclusions}\label{sec:con}
In this paper we have faced a fundamental question intrinsically related with
feedback Brownian ratchets, namely, the effects of a time delay in such a
feedback controlled stochastic system. We have focused on the
task of studying the dependence of the flux with the time delay
for both the case of one particle and for the collective version
of the ratchet with few particles.
\par

For \emph{one particle} ratchets and small delays we have obtained
an effective potential which contains the basic ingredients that
come into play, and gives an approximate analytical expression for
the flux. The effects of the delay in the shape and the average
slant of the effective potential allows one to easily understand the
decrease of the flux with increasing delays. The approximate
analytical expression obtained [Eqs.~\eqref{FPE-sol} and
\eqref{pot-sol}] gives the average velocity in terms of the main
magnitudes of the system: the height of the potential $V_0$, its
asymmetry $a$, and the time delay in the feedback $\tau$. In
particular, it allows one to obtain predictions of the characteristic
time scale of the decrease due to the delay. This relation is also
useful in the \emph{few particle} case thanks to the relation
\eqref{v-few} found between the flux for the one and the few particle
cases.
\par

The decrease of the covariance of the sign of the net force for
increasing delays provides an alternative approach to understand
the dependence of the flux with the delay. This approach has given
the relation between the covariance and the flux, and has allowed us
to relate the flux obtained in the few particle case with the
results of the one particle case [Eq.~\eqref{v-few}]. In addition,
the fact that the covariance becomes negligible for large delays
indicates that the delayed control protocol effectively behaves as
if it were an open-loop control protocol. This results in a
constant value of the flux for large delays that is independent on
the number of particles.
\par

We want to stress as an important result of this paper that the
feedback controlled system for one or few particles is able to
perform better than its open-loop counterpart even for nonzero
time delays (provided the delays are smaller than the
characteristic times of the dynamics of the Brownian ratchet).
Furthermore, even for arbitrarily large delays the net flux does
not vanish but it reaches a positive value, albeit it performs
worse than the optimal open-loop protocol. We also highlight the
importance of this study for realistic \emph{experimental}
situations that necessarily have to face with time delays. For
the ratchet considered in~\cite{rou94} the colloidal particles
have diameter $0.25$, $0.4$, and
$1\;\mu\text{m}$, and the sawtooth dielectric potential has period
$L=50\;\mu\text{m}$ and asymmetry $a\sim 1/3$. The maximum
velocities obtained with a periodic switching were
reported~\cite{rou94} to be of $0.2\;\mu\text{m/s}$ with ${\cal
T}_{\text{on}}\sim 30\;\text{s}$ and ${\cal T}_{\text{off}} \sim
50 \;\text{s}$. As the trapping energy is significantly greater
than $kT$ and $a\sim 1/3$ the introduction of feedback can
increase the velocity up to a factor $(1/2-a)^{-1}\sim 6$
approximately~\cite{cao04,bie07}, attained when the time delay in
the feedback is negligible. The results obtained in this paper
indicate that for delays in the feedback smaller than the
characteristic times of the system (of order $10\;\text{s}$,
\emph{i.e.}, of order $10^{-3}$ in the adimensional units used
throughout our paper) it is possible to obtain velocities greater
than the maximum of open-loop protocols. The use of a conventional
CCD camera ($30\;\text{fps}$) and conventional electronics is
enough to achieve a feedback control performance with a time delay
of the order of $0.1\;\text{s}$ ($10^{-4}$ in adimensional units),
for this time delay an increase of the velocity of a factor of $4$ is
expected. This points towards the feasibility of implementing
experimentally a feedback controlled ratchet that performs better
than its optimal open-loop version.
\par

We have also studied the effects of time delays in the \emph{many
particle} case, where surprising and interesting results arise.
Although in the many particle case without delay the instantaneous
maximization protocol performs worse than the optimal open-loop
protocol, the introduction of a delay can increase the
center-of-mass velocity up to the values given by the optimal
open-loop control protocol. For small delays the asymptotic
average velocity decreases for increasing delays, until it reaches
a minimum. After this minimum, a change of regime happens and the
system enters a selfsynchronized dynamics with the net force at
present highly correlated with the delayed value of the net force
used by the controller. This self-synchronization stabilizes
several of the quasiperiodic solutions that can fit an integer
number of periods in a time interval of the length of the time
delay. The stable quasiperiodic solutions have a structure
similar to those solutions appearing in the threshold protocol.
This analogy has allowed us to make numerical and analytical
predictions using the previous results for the threshold protocol
\cite{fei06}. In particular, we have established the location and
value of the maximum, and also the value of the time delay beyond
which a quasiperiodic solution becomes unstable. The results
obtained show that for most time delays several solutions are
stable and therefore the systems present multistability; which
stable solution is reached depends on the past history of the
system. The possibility to choose the quasiperiod of the solution
we want to stabilize just tuning the time delay can have potential
applications to easily control the particle flux. Note that we can
even leave some branch just going to time delays where the branch
is already unstable, and force the system to change to another
branch of solutions.
\par

In summary, we have studied the effects of time delays in the feedback control
of a flashing ratchet. The results for one and few particles point towards the
feasibility of an experimental implementation of a feedback controlled ratchet
that performs better than its optimal open-loop version. On the other hand, the
many particle case presents an unexpected improvement of the
flux due to the stabilization of one or more quasiperiodic solutions for large
enough delays.

\acknowledgments

We acknowledge financial support from the MEC (Spain) through
Research Projects FIS2005-24376-E and FIS2006-05895, and from the
ESF Programme STOCHDYN. In addition, M.F. thanks the Universidad
Complutense de Madrid (Spain) for support through grant ``Beca
Complutense''.


\begin{thebibliography}{0}
\bibitem{smo12} M. V. Smoluchowski, Phys. Z. 13, 1069 (1912).
\bibitem{fey63} R. P. Feynman, R. B. Leighton, and M. Sands, The Feynman
  Lectures on Physics (Addison-Wesley, Reading, MA, 1963).
\bibitem{ajd93} A. Ajdari and J. Prost,
  C. R. Acad. Sci. Paris II {\bf 315}, 1635 (1993).
\bibitem{mag93} M. O. Magnasco, Phys. Rev. Lett. {\bf 71}, 1477 (1993).
\bibitem{ast94} R. D. Astumian and M. Bier, Phys. Rev. Lett. {\bf 72}, 1766
  (1994).
\bibitem{rei02} P. Reimann, Phys. Rep. {\bf 361}, 57 (2002).
\bibitem{lin02} H. Linke, Appl. Phys. A {\bf 75}, 167 (2002).
\bibitem{cao04} F. J. Cao, L. Dinis and J. M. R. Parrondo,
  Phys. Rev. Lett. {\bf 93}, 040603 (2004).
\bibitem{din05} L. Dinis, J. M. R. Parrondo, and F. J. Cao,
  Europhys. Lett. {\bf 71}, 536 (2005).
\bibitem{fei06} M. Feito and F. J. Cao, Phys. Rev. E {\bf 74}, 041109 (2006).
\bibitem{rou94} J. Rousselet, L. Salome, A. Ajdari, and J. Prost, Nature {\bf
    370}, 446 (1994).
\bibitem{mar02} C. Marquet, A. Buguin, L. Talini, and P. Silberzan,
  Phys. Rev. Lett. {\bf 88}, 168301 (2002).
\bibitem{bie07} M. Bier, Biosystems {\bf 88}, 301 (2007).
\bibitem{ste94} R. F. Stengel, \textit{Optimal Control and Estimation} (Dover,
  New York, 1994).
\bibitem{bec05} J. Bechhoefer, Rev. Mod. Phys. \textbf{77}, 783 (2005).
\bibitem{boc00} G. A. Bochanov and F. A. Rihan, J. Comput. Appl. Math. {\bf
    125}, 183 (2000).
\bibitem{fra05b} T. D. Frank,
  Phys. Rev. E {\bf 71}, 031106 (2005).
\bibitem{kos05} M. Kostur, P. H\"anggi, P. Talkner, and J. L. Mateos,
  Phys. Rev. E {\bf 72}, 036210 (2005).
\bibitem{son06} W. S. Son, J. W. Ryu, D. U. Hwang, S. Y. Lee, Y. J. Park,
  C. M. Kim, nlin.CD/0612039 preprint (2006).
\bibitem{gui99} S. Guillouzic, I. L'Heureux and A. Longtin,
  Phys. Rev. E {\bf 59}, 3970 (1999).
\bibitem{fra05} T. D. Frank, Phys. Rev. E {\bf 72}, 011112 (2005).
\bibitem{fra02} T. D. Frank, Phys. Rev. E {\bf 66}, 011914 (2002).
\bibitem{ris89} H. Risken, \emph{The Fokker-Planck equation, Methods of
    Solution and Applications} (Springer-Verlag, Berlin, 1989).
\bibitem{abra} M. Abramowitz and I. A. Stegun, \emph{Handbook of Mathematical
    Functions} (Dover, New York, 1972).
\bibitem{cao07} F. J. Cao, M. Feito and H. Touchette, \emph{Information
  and flux in a feedback controlled Brownian ratchet},
  arXiv:cond-mat/0703492 (2007).
\bibitem{fei07} M. Feito and F. J. Cao, Eur. Phys. J. B {\bf 59}, 63 (2007).
\end{thebibliography}
\end{document}